\documentstyle[epsf,twoside,fleqn,mystyle]{article}

\title{On Gauge Invariance and Ward Identities for the Wilsonian 
Renormalisation Group
\thanks{Talk presented by JMP at QCD 98, Montpellier, 
July 1998. To be published in the proceedings.}
}

\author{
Daniel F. Litim${}^{a}$ and Jan M. Pawlowski${}^{b}$\\[2ex]
${}^a$Departament ECM \& IFAE,
Facultat de F\'{\i}sica, Universitat de Barcelona\\
$\, \;$Diagonal 647, E-08028 Barcelona, Spain.\\[2ex]
${}^b$Dublin Institute for Advanced Studies,
10 Burlington Road, Dublin 4, Ireland.\\[2ex]}
\begin{document}
\thispagestyle{empty}

\begin{abstract}
We investigate non-Abelian gauge theories within a Wilsonian Renormalisation 
Group approach. The 
cut-off term inherent in this approach leads to a modified Ward identity 
(mWI). It is shown that this mWI is compatible with the flow and that the full 
effective action satisfies the usual Ward identity (WI). The universal 1-loop 
$\beta$-function is derived within this approach and the extension to the 
2-loop level is briefly outlined. 
\vspace*{-8.2cm}\begin{flushright}{\normalsize DIAS-STP-98-08\\ 
ECM-UB-PF-98-20}\end{flushright}
\vspace*{6.7cm}
\end{abstract}

\maketitle

\section{Introduction}
The Wilsonian Renormalisation Group \cite{Wilson} has proven itself as a 
powerful tool for studying both perturbative and non-perturbative effects 
in quantum field theory. 
One may expect that a suitable formulation for non-Abelian 
gauge theories provides new insight to non-perturbative effects in QCD. 
However the Wilsonian approach is based on the concept of a step-by-step 
integrating-out of momentum degrees of freedom and one may wonder whether this 
concept can be adopted for gauge theories. 

In the current contribution we investigate this question in a path integral 
approach based on ideas of Polchinski \cite{ERG}. In this approach 
a momentum cut-off is achieved by adding a cut-off term $\Delta_k S$ 
to the action which 
is quadratic in the field. This results in
 an effective action $\Gamma_{k}$ where momenta larger 
than $k$ have been 
integrated-out. The change of $\Gamma_k$ under 
an infinitesimal variation of the scale $k$ is described by a flow equation
which can be used to successively integrate-out the momenta 
smaller than the cut-off scale $k$. Thus given an effective action 
$\Gamma_{k_0}$ at an initial scale $k_0$ the flow equation provides us with a 
recipe how to calculate the full effective action $\Gamma$.  

The introduction of $\Delta_k S$ seems to 
break gauge invariance. However, $\Gamma_k$ satisfies a `modified' 
Ward identity (mWI). This mWI commutes with the flow and 
approaches the usual Ward identity (WI) as $k\to 0$. 
Consequently the full effective action $\Gamma$ satisfies the usual Ward 
identity. In other words, gauge invariance of the full theory is preserved 
if the effective action $\Gamma_{k_0}$ satisfies the mWI 
at the initial scale $k_0$. 

\section{The Flow Equation}
To be more explicit let us 
briefly outline the derivation of the flow equation: We add the following
 scale-dependent term to the action (e.g.\ \cite{collect}, \cite{gag} and 
references therein): 
\bg\label{cut-off1} 
\Delta_k S[\Phi] = \frac{1}{2} \int \d^d x\,\Phi^*
R^{\Phi}_k[P_\Phi]\Phi,
\eg 
where $P_\Phi^{-1}$ is proportional to 
the bare propagator of $\Phi$, $(\Phi_i)=(\Phi_1,...,\Phi_n)$ 
is a short-hand notation for all fields. $d$ is the 
dimension of space-time. The regulator $R_k^\Phi$ has the properties: 
\bg \label{limits}
R_k^{\Phi}[x]\stackrel{{x\ov k^2}\rightarrow 0}{\longrightarrow} 
k^{d-2 d_\Phi} \frac{x}{|x|}, \ \ \ \  
R_k^\Phi[x]
\stackrel{{x\ov k^2}\rightarrow \infty}{\longrightarrow}  0,
\eg 
where $d_\Phi$ are the dimensions of the fields $\Phi$. The 
cut-off term \eq{cut-off1} 
effectively suppresses modes with momenta  
$p^2\ll k^2$ in the generating functional. 
For modes with large momenta $p^2\gg k^2$ the cut-off term 
vanishes and in this regime the theory remains unchanged. In the limit 
$k\rightarrow 0$ we approach the full generating functional $\Gamma$ since 
the cut-off term is removed. In the limit $k\to \infty$ all momenta are 
suppressed and the effective action approaches the (gauge fixed) 
classical action $S_{cl}+S_{gf}$. 
Hence $\Gamma_k$ interpolates between the classical action and 
the full effective action: 
\bg\label{0infty}
S_{cl}+S_{gf}\stackrel{k\to\infty}{\longleftarrow}
\Gamma_k \stackrel{k\to 0}{\longrightarrow} \Gamma.
\eg 
An infinitesimal 
variation of the generating functional with respect to
 $k$ is described by the flow equation. 
For the generating functional of 1PI Green functions, 
the effective action $\Gamma_k$, the flow equation 
can be written in the form (e.g. \cite{collect}, \cite{gag} and references 
therein)
\bg\label{flow}
\partial_t\Gamma_k[\Phi]=  \frac{1}{2}\STr\left\{ G^{\Phi^*\Phi}_k[\Phi]
\partial_t R^\Phi_k[P_\Phi]\right\},
\eg 
with 
\bg\label{props}
G^{\Phi^*_i \Phi_j}_k[\Phi]=  
\left(\frac{\delta^2\Gamma_k[\Phi]}{\delta\Phi^*_i \delta\Phi_j} + 
R_{k,ij}^{\Phi}[P_\Phi]\right)^{-1},
\eg 
where $t=\ln k$ and 
the trace $\STr$ denotes 
a sum over momenta, indices and the different fields $\Phi$ including a minus 
sign for fermionic degrees of freedom. 
Note that $\partial_t R_k^\Phi$ 
serves as a smeared-out $\delta$-function in momentum space peaked at about 
$p^2\approx k^2$. Thus by varying the scale $k$ towards smaller $k$ 
according to \eq{flow} one successively integrates-out 
momentum degrees of freedom. 

\section{QCD in a Wilsonian approach}
The starting point of our considerations is the classical action of QCD with 
$N_c$ colours and $N_f$ flavours 
including cut-off terms for the gauge field and the fermions: 
\bg\no
S_k[A,\psi,\bar\psi] &\!\! = &\di\!\!S_{cl}[A,\psi,\bar\psi]+S_{gf}[A]\\\di 
&&\di+\Delta_k S_A[A]+\Delta_k S_\psi[\psi,\bar\psi],
\label{classical}\eg 
where $S_{cl}$ is just the classical Euclidean action of QCD with fermions 
in the fundamental representation. 
We allow for a general linear gauge fixing term 
\bg\label{gf}
S_{gf}[A]=\frac{1}{2\xi}\int \d^4 x\, l_\mu A_\mu^a\, l_\mu A_\mu^a, 
\eg 
which includes general Lorentz gauges ($l_\mu=\partial_\mu$) 
and general axial gauges 
(e.g. \cite{gag},\cite{we}). The cut-off terms are given by  
\bg\no
\Delta_k S_A[A]&\!\! =\tab \!\!
\frac{1}{2}\int \d^4 x\, A_\mu^a R_{k,\mu\nu}^{A,ab} A_\nu^b,\\\di 
\Delta_k S_\psi[\psi,\bar\psi]&\!\!
=\tab \!\!\int \d^4 x\, \bar\psi_s^A 
R^{\psi,AB}_{k,st} \psi_t^B, 
\label{cut-off}\eg 
where $A,B$ and $a,b$ refer to the fundamental and to the 
 adjoint representations respectively. The indices $s,t$ are summed over all 
flavours. A convenient choice for the regulators $R^A_k, R^\psi_k$ is 
\bg\no
 R_{k,\mu\nu}^{A,ab}[p^2]&\!\!=\tab \!\!
\delta_{\mu\nu} \delta^{ab}\frac{p^2}{e^{p^2/k^2}-1},
\\\di 
 \hat R^{\psi,AB}_{k,st}[p^2]&\!\!=\tab \!\!\delta_{st} \delta^{AB}\left[
\frac{p^2}{e^{p^2/k^2}-1}\right]^{1\ov 2}.
\label{regulators}\eg 
It is easy to see that the regulators in \eq{regulators} have the demanded 
properties \eq{limits}. 

The Fadeev-Popov determinant arising from the 
gauge fixing \eq{gf} may be regularised in a similar way. However, 
for the sake of brevity we drop these terms in the following. 

The cut-off terms \eq{cut-off} generate additional terms in the Ward identity 
for $\Gamma_k$. This modified Ward identity (mWI) is 
\bg\label{mwi}\no
\lefteqn{\CW_k^a(x)= 
D_\mu^{ab}\frac{\delta\Gamma_k}{\delta 
A^b_\mu}-D_\mu^{ab}{l^*_\mu l_\nu \ov \xi}
A_\nu^b+J^{\psi,a}}\\\no\di
&&\di -g \int \d^4 y\, f^{abc}\left(\frac{l^*_\mu
  l_\nu}{\xi}\delta^{cd} +R^{A,cd}_{k,\mu\nu}\right)
  G^{AA,db}_{k,\nu\mu}\\\di 
&&\di +g \int\d^4 y\, (t^{a})^{BC}R^{\psi,CD}_{k,st}
 G^{\psi\bar\psi,DB}_{k,ts} =0 
\eg 
where $t^a$ are the gauge group generators in the fundamental representation 
and $l_\mu^*$ is the adjoint of $\l_\mu$. 
We also have used the definition of the full (field dependent) 
propagators \eq{props} and 
have introduced the following short-hand notation: 
\bg\label{jpsi}
J^{\psi,a}=\bar\psi^A(t^a)^{AB} {\delta\Gamma_k\ov \delta \bar\psi^B}
+{\delta 
\Gamma_k\ov \delta \psi^A}(t^a)^{AB}\psi^B. 
%G^{AA}_{k}=\tab\!\!
%\left(\frac{\delta\Gamma_k}{(\delta A)^2 
%} +R_{k}^{A}\right)^{-1},\\\di  
%G^{\psi\bar\psi}_{k}
%=\tab\!\!\left(\frac{\delta}{\delta \psi\bar\psi}\Gamma_k+
%R_{k}^{\psi}\right)^{-1}. 
\eg 
The cut-off dependent terms in \eq{mwi} vanish for $k\to 0$. However 
the mWI \eq{mwi} is of use only if one can show 
that the flow of $\Gamma_k$ is compatible with \eq{mwi}. For this purpose
 we examine $\partial_t \CW_k^a$. The $t$-derivative of the right-hand 
side of \eq{mwi} yields expressions dependent 
on $\partial_t R^A_k,\partial_t 
R^\psi_k$ and $\partial_t \Gamma_k$. For the last of these we use the flow 
equation \eq{flow} and after some algebra we arrive at [$\Phi=
(A,\psi,\bar\psi),\ \Phi^*=(A,-\bar\psi,\psi)$]: 
\bg\no
\lefteqn{\partial_t \CW^a_k=} \\\di 
&&\di \!\!\!\!\!\! -\frac{1}{2}\STr\left( G^{\Phi^*_i\Phi_j}_k \partial_t 
R^{\Phi_j}_k G^{\Phi^*_j\Phi_l}_k \frac{\delta}{\delta \Phi^*_l}
\frac{\delta}{\delta \Phi_i}\right)\CW^a_k. 
\label{compatible}\eg 
Let us assume that the initial effective 
action $\Gamma_{k_0}$ satisfies the mWI which can be achieved at 
least order-by-order in perturbation theory. 
Then $\partial_k\CW_{k}|_{k_0}$ is zero since it is proportional to 
$\CW_{k_0}$. Thus it follows $\CW_k^a=0$ for all $k$ and the mWI
 is satisfied at all scales, in particular for $k=0$. It  
 approaches the usual WI for $k\to 0$ since the cut-off dependent 
terms vanish. As a consequence we only have to ensure that the initial 
effective action satisfies the mWI in order to ensure gauge 
invariance of the full theory. 

\section{Applications}
As a first application we want to present some analytic results. 
Analytic computations would simplify tremendously if we still dealt with 
a theory satisfying the usual WI instead of \eq{mwi}. In this case the number 
of possible terms in the effective action is restricted by gauge 
symmetry. Even though this cannot 
be achieved one can get very close to such a situation. For this 
purpose it is quite convenient to introduce the following regulators:  
\bg 
(R^A_k[p^2],R^\psi_k[p^2])\!\to\!(R_k[D_T(\bar A)],R^\psi_k[\dr^2(\bar A)]),
\label{newregs}\eg 
with 
\bg 
D_{T,\mu\nu}^{ab}=-D_\rho^{ac}
D_\rho^{cb} -2 g f_c^{ab}F_{\mu\nu}^c, 
\eg 
and $\dr$ is the Dirac operator in the fundamental representation. 
Here $\bar A$ is an arbitrary gauge field configuration. Note that if one 
allows for regulators with a non-trivial group structure the set 
of regulators $R^A_k[D_T],R^\psi_k[\dr^2]$ 
coincides with the set of regulators for $R^A_k[p],R^\psi_k[p]$: 
$\{ R^A_k[p],R^\psi_k[p]\}=\{R_k[D_T],R^\psi_k[\dr^2]\}$. Thus one may 
interpret $\bar A$ as an index labeling a family of different cut-offs. 
The effective action now depends on $\bar A$: $\Gamma_k=
\Gamma_k[A,\bar A]$. 

It is simple to see that the cut-off dependent terms in 
\eq{mwi} are just given by an infinitesimal gauge transformation of $\bar A$, 
hence leading to the identity 
\bg\no
\lefteqn{D_\mu^{ab}(\bar A)\frac{\delta\Gamma_k}{\delta 
\bar A^b_\mu}=} 
\\\no\di & &\di\!\! 
-g \int\d^4 y\, f^{abc} R^{A,cd}_{k,\mu\nu}[D_T(\bar A)] 
G^{AA,db}_{k,\nu\mu}\\\di 
&&\di\!\!\!\!\! +g \int\d^4 y\, (t^{a})^{BC}R^{\psi,CD}_{k,st}[\dr^2(\bar A)]
 G^{\psi\bar\psi,DB}_{k,ts}\label{barwi}.  
\eg 
We conclude from \eq{mwi} and \eq{barwi} that $\hat\Gamma[A,\psi,\bar\psi]:=
\Gamma_k[A,\bar A=A,\psi,\bar\psi]$ satisfies the 
usual WI without the cut-off dependent terms: 
\bg\no
D_\mu^{ab}\frac{\delta\hat\Gamma_k}{\delta 
A^b_\mu}-D_\mu^{ab}{l^*_\mu l_\nu \ov \xi}
A_\nu^b+J^{\psi,a}(x)\\\di
-g \int\d^4 y\, f^{abc}\frac{l^*_\mu
  l_\nu}{\xi}\delta^{cd} 
  G^{AA,db}_{k,\nu\mu} =0,\label{wi} 
\eg 
where the gauge field derivative involved in \eq{wi} hits both the 
gauge field $A$ and the auxiliary field $\bar A=A$. Note however that 
the propagator $G^{AA}_k$ is still the one derived from 
${\delta^2\ov \delta A^2}\Gamma_k[A,\bar A]$ at $\bar A=A$. Moreover 
the flow 
equation for $\hat\Gamma_k$ requires the knowledge of $G^{AA}_k$, thus 
slightly spoiling the advantage of dealing with an effective action 
which satisfies the usual WI even for $k\neq 0$. 

A possible way to 
proceed from \eq{wi} would be to reformulate it
 in terms of BRST transformations of the fields. Then one can 
expand the action $\hat\Gamma_k$ in BRST-invariant terms. 

Moreover if we 
restrict ourselves to gauges where neither $\xi$ nor $\l_\mu$ depend on 
derivatives, the last term in \eq{wi} vanishes 
and we do not have any integral terms in the WI (see also \cite{we}). 
This is a very attractive case where all the following considerations 
simplify tremendously from a technical point of view.  

As a consequence of \eq{wi} 
we have gained gauge invariance even for $k\neq 0$ which 
simplifies the expansion of the effective action. The 
problem is now to distinguish between the gauge field $A$ and the 
field $\bar A=A$ which only serves as an auxiliary variable. This is 
necessary since the flow equation still requires the knowledge of $G_k^{AA}$ 
as mentioned above. The $\bar A$-dependence of 
$\partial_t\Gamma_k$ is given by the following equation:
\bg
\frac{\delta }{\delta \bar A}\partial_t\Gamma_k=
\frac{1}{2}\partial_t
\STr \left\{G^{\Phi^*\Phi}_k[\Phi,\bar A]{\delta\ov \delta 
\bar A} R^{\Phi}_k\right\}. 
\label{bar A}\eg 
With \eq{flow}, \eq{mwi}, \eq{barwi} and \eq{bar A} we can 
investigate the effective action analytically. It is worth noting that the 
flow equation is a `1-loop' equation, even though the loops depend on the full 
field dependent propagator. Thus heat kernel methods can be employed. 
However we want to emphasise that the 
heat kernel is not used as a regularisation method since everything is finite 
from the onset. 

Let us now briefly sketch the calculation of the (perturbative) 1-loop and 
2-loop $\beta$-function. Since these coefficients of the 
$\beta$-function are universal their calculation 
 serves as a consistency check of the formalism. Moreover it provides some 
additional insight in how perturbation theory is recovered in this approach. 

On the right-hand side of the flow equation \eq{flow} we 
have to insert the initial effective action $\Gamma_{k_0}$. 
At 1-loop level it is sufficient to insert 
the classical action $S_{cl}+S_{gf}$ 
\eq{classical} with multiplicative 
renormalisation, namely $A\to Z^{1/2}_{F,k} A,\ g\to 
g_k=Z_{g,k} g$. Moreover at 1-loop level one can show with the help of 
\eq{wi} and \eq{bar A} 
that the usual relation between $Z_{g,k}$ and $Z_{F,k}$ 
is valid. This relation 
depends on the chosen gauge, e.g.\ for the axial gauge it 
leads to $\partial_t Z^2_{g,k}/Z^2_{g,k} =-\partial_t Z_{F,k}/Z_{F,k}$ 
(see also \cite{we}). $\partial_t Z_{F,k}/Z_{F,k}$ is calculated by 
projecting out the term $\tr\, F^2$ on the right hand-side side of 
the flow equation \eq{flow}. The result leads to 
($\beta_{g^2}=\partial_t g_k^2$): 
\bg
\beta_{g^2}=-\frac{1}{16\pi^2}g_k^4\left({22\ov 3}N_c-{4\ov 3} N_f\right)
+O(g_k^6), 
\eg 
the well-known universal 1-loop result for a non-Abelian gauge theory 
coupled to fermions in the fundamental representation. Moreover it
 can be shown that this holds true for general linear gauges. 

In order to derive the 2-loop coefficient for the $\beta$-function one has 
to take into account not only the renormalisation constants $Z_F,Z_g$ but also 
terms which can be derived from the mWI \eq{mwi} when 
examined at 1-loop level. Additionally one has to examine \eq{bar A} which in 
general is non-zero at 2-loop level \cite{2-loop}. 

We would like to emphasise that the calculations outlined above not only 
provide the $\beta$-functions but also furnish one with correction terms 
to the effective action at 2-loop level
to all orders of the fields. 

\section{Conclusions}
We have investigated non-Abelian gauge theories coupled to fermions 
within a Wilsonian renormalisation group  
approach. Gauge invariance of the effective action at a given (infrared) 
scale $k$ is controlled by a modified Ward identity which is 
compatible with the flow equation. The mWI guarantees gauge invariance for the 
full effective action at $k=0$. By introducing an auxiliary gauge field 
$\bar A$ gauge invariance can be restored even for $k\neq 0$. The price to 
pay is an additional equation for the $\bar A$-dependence. 

As a consistency check the 1-loop 
$\beta$-function can be calculated for general linear gauges. Even more so, 
the extension of this calculation to 2-loop effects is straightforward 
(even though tedious) \cite{2-loop}. 

The calculations presented here also give a flavour of the main 
advantage of the formalism, namely its flexibility concerning possible 
approximations, in particular beyond perturbation theory. 
This makes it an appropriate tool for studying 
non-perturbative physics.

\end{document}